# PHOTOMETRIC DATA ANALYSIS OF THE ECLIPSING BINARY SYSTEM AH TAURI

**M. A. El-Sadek*, N. S. Awadalla*, Ahmed-Essam* and M. A. Rassem****

**ABSTRACT:** ***Two sets of photometric observations of the system AH Tauri have been analyzed using the latest version of the Wilson-Devinney code. The results show that AH Tauri may classified as A-type of W-UMa eclipsing binary. The mass ratio of q = 0.81, an over-contact degree of f=0.095, and a slightly temperature difference between the two components have been obtained. The asymmetry of its light curve explained by the presence of a dark spot on the massive component. The physical, geometrical, and absolute parameters have been derived and compared with previous work.***

## INTRODUCTION

AH Tauri (2MASS J03471196+2506593, GGM2006 6734014, HV 6187, CSI +24-3442, $\alpha(2000) = 03^h\ 47^m\ 11^s.964$, $\delta(2000) = +25°\ 06'\ 59''.31$) is a W- UMa eclipsing binary system, with a magnitude $V_{max.}$ = 11.25 and $V_{min.}$ = 11.92. It is a short orbital period system (P=0.33267174 day). So, possibly, its light curve can be observed completely during one night. The photometric variability of the system has been discovered by Shapley et al. (1934). Its photographic observations have been found by Binnendijk (1950) and Romano (1962). Binnendijk (1950) had classified the system as W- UMa type eclipsing binary with spectral type of G1p, while Romano (1962) had classified it as β Lyr type. The first photoelectric observations in B and V bands have been published by Bookmyer (1971) with partial covering of phase. Bookmyer (1971) indicated that the spectral type of AH Tau is about G5 and its color index (B-V) of 0.66. The first complete photoelectric light curves in B and V bands have been obtained by Magalashvili et al. (1980). They noticed that the light curves were asymmetrical and the secondary mid-eclipse occurred at a phase of 0.47. Liu et al. (1991) have been deduced a photometric solution from new photoelectric observations in B and V bands using the Wilson-Devinney code. Their observations have been

**** National Research Institute of Astronomy and Geophysics, Helwan, Cairo, Egypt***
***** Astronomy Department, Cairo University, Egypt***

showed some differences in the light curves from those obtained by Magalashvili et al. (1980), who classified the system AH Tau as a typical W- UMa binary with an overcontact of 9.1%. The time of minima for AH Tau were published by Binnendijk (1950), Romano (1962), Bookmyer (1971), Magalashvili et al. (1980), Liu et al. (1991), Nelson (2001), and Pribulla (2001). Yang and Liu (2002) showed that the changes of the orbital period of AH Tau continuously decreased at a rate of $\Delta p/p = -1.4*10^{-11}$ from 1944 to 1976, then suddenly and sharply decreased by about 0.33 s around 1976. After 1976, the orbital period of the system continually increased at a change rate of $\Delta p/p = 1.5*10^{-10}$. This may be due to the compound result of three mechanisms: the cyclical magnetic activity, the mass loss and the radius swelling of the two components of the system (Yang and Liu (2002)).

## PHOTOMETRIC DATA ANALYSIS

Two sets of non-analyzed observational data have been collected for the contact binary system AH Tau, First set is a photographic observation, which have been done in the year 1943 using the 33-cm Leiden photographic refractor and published as normal points by Binnendijk (1950). The mean error of the normal points is found to be $\pm 0.011_{mag}$. Second set was observed in the night of Nov 29, 2002 and published on Dec 1, 2002 in the web site by Nakajima (2002). It was a photometric CCD observation in V band.

Figures 1 & 2 represent the observational normal points (Phase & Mag.) in B band for the observational data of 1950 and in V band for the observational data of 2002 respectively, by using a new light elements for the AH Tau. system as;

$$\text{Hel. J.D. (Min.I)} = 2452607.9552 + 0.33267174 * E$$

The analyses of the photometric data in B band of Binnendijk (1950), and in V band of Nakajima (2002) have been carried out using Wilson-Devinney (WD) program (Wilson et al., 2003). The model has been described and quantified in papers by Wilson & Devinney (1971), Wilson (1979, 1990, 1993), and Van Hamme & Wilson (2003).

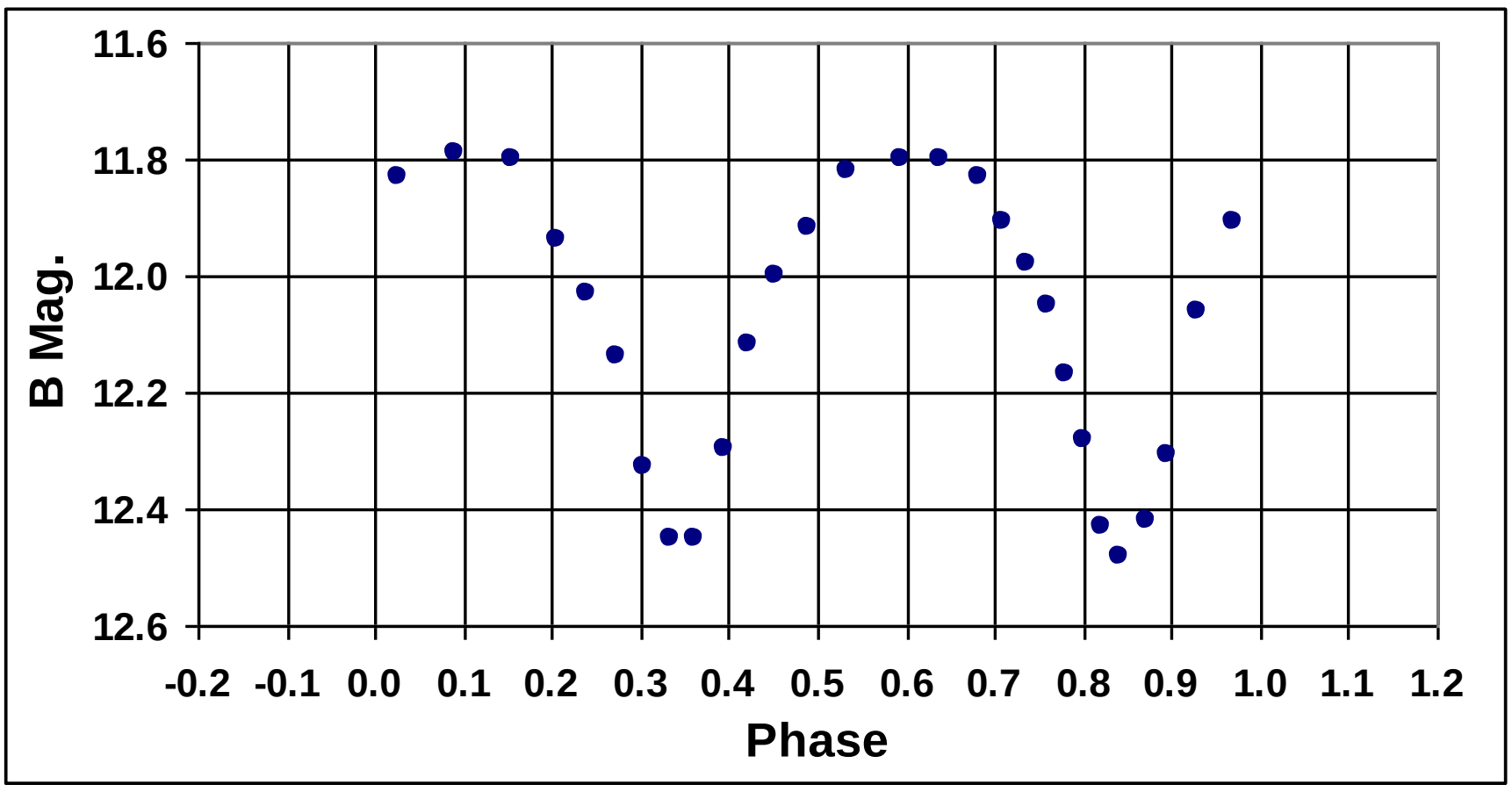

**Fig. 1: The normal points B light curve of AH Tau., on 1950**

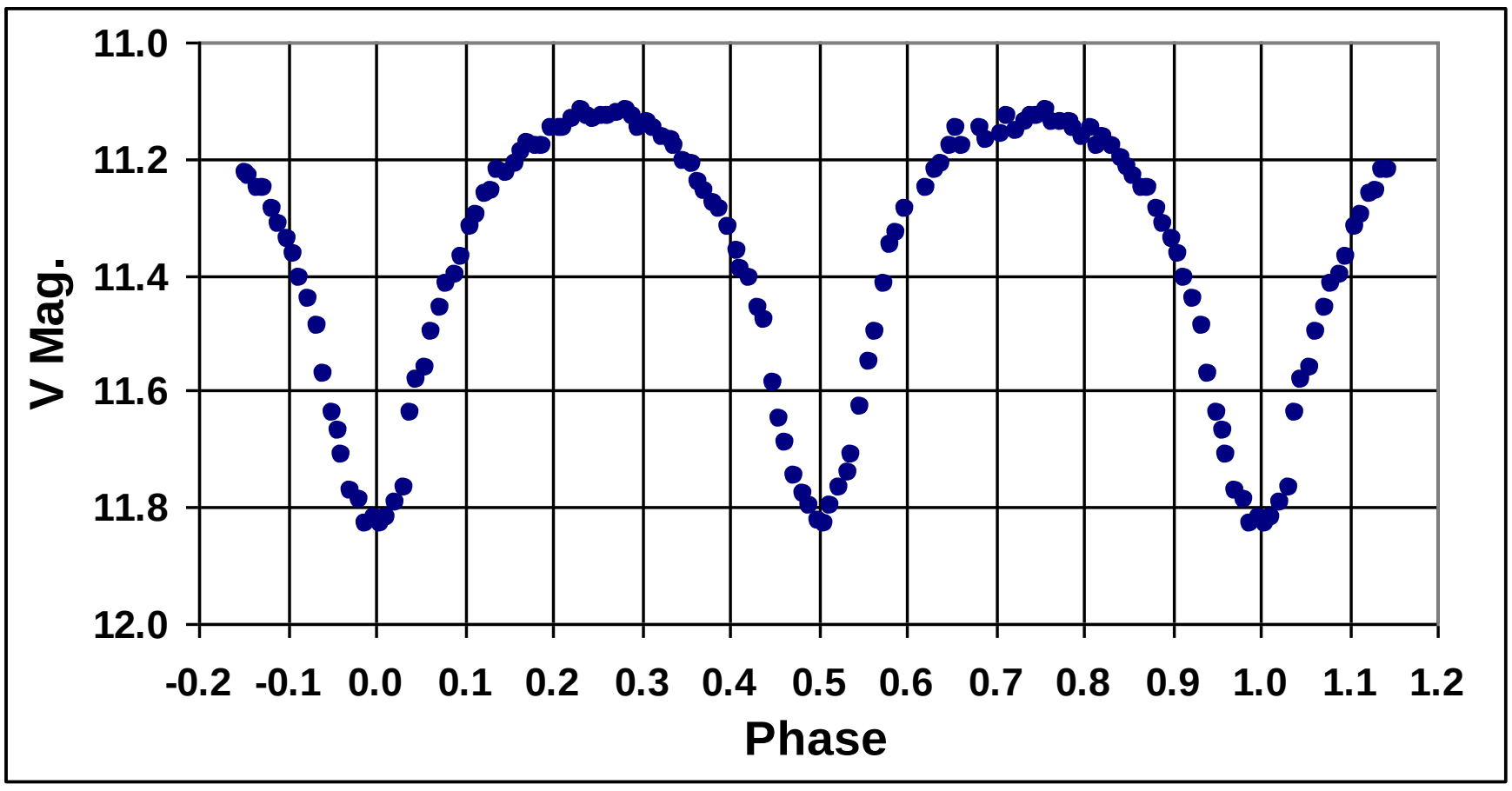

**Fig. 2: The normal points V light curve of AH Tau., on 2002**

The exponent of gravity darkening, $g_1 = g_2 = 0.32$ (Lucy 1967) and bolometric albedo, $A_1 = A_2 = 0.5$ (Rucinski 1969) for late spectral type stars have been assumed. The values of bolometric limb darkening, $x_1$ and $x_2$, have been taken from Van Hamme Tables (1993). A cool spot on star 1 has been supposed to treat the asymmetry light curve especially in the 2002 Nakajima's observations. The overcontact mode 3 of WD 2003 code has been used too. The orbital inclination i, the mass ratio q, and some spot parameters of binary component have been adopted. The parameters; separation between the two components, a, the mean surface temperature of the stars 1 and 2, ($T_1$, $T_2$), potential of two components, ($\Omega_1 = \Omega_2$), monochromatic limb darkening ($x_1$ & $x_2$), and the monochromatic luminosity of star 1 ($L_1$) have been adjusted and employed in the second part of WD programme (Differential Correction programme; DC). The relative brightness of the star 2 was calculated by the stellar atmosphere model.

The orbital, physical, and absolute parameters of the AH Tau system are presented in Table 1. The obtained results have been compared with the results of Byboth et al (2004) and Liu, et al. (1991), where star 1 indicates the massive and hot component and star 2 the less massive and cooler one. The final fit of the observations have been represented in figure 3 for B band of 1950 observations and in figure 4 for V band of 2002 observations. The fitting results are believed to be quite satisfactory.

To draw the roche geometry outline and to show the degree of contact, the Binary Maker 3 programme (Bradstreet, 2004) has been used. The obtained results from the BM3 have been represented in figure 5 for B band of 1950 observations and in figure 6 for V band of 2002 observations. In addition, the 3 dimension model of the 1950 observations of AH Tau. for phases 0.0 and 0.5 has been represented in figure 7 without spot in B filter, while the position of the spot has been represented at phases 0.0 and 0.5 in figure 8 for V band of 2002 observations.

The obtained results for AH Tau have been used too to classify the system either A- or W- subtype of W-UMa binary stars. Table 2 shows the light curves change of AH Tau. While Table 3 shows the differences between sub-type A and subtype W for the contact binary AH Tau.

**Table 1: Comparison between different sets of observation ns**

| Parameters | Present Analysis Binnendijk, 1950 B band | Present Analysis Nakajima, 2002 V band | Byboth, 2004 V band | Liu, et al. 1991 V band |
|---|---|---|---|---|
| Wavelength = λ | 4330 A° | 5500 A° | 5500 A° | 5500 A° |
| Phase Shift | -0.1533 ± 0.0014 | 0.0004 ± 0.000 | 0.000 | 0.000 |
| Inclination (i) | 78°.00 (assumed) | 80°.70 fixed | 80°.73 ± 0°.12 | 84°.29 ± 0°.19 |
| Surface Temp. $T_1$<br>Surface Temp. $T_2$ | 5897 ºk ± 26.7 ºk<br>5873 ºk ± 26.4 ºk | 5885 ºk fixed<br>5852 ºk ± 9.4ºk | 5900 ºk (fixed)<br>5815 ºk ± 11 ºk | 5890 ºk (fixed)<br>5866 ºk ± 5 ºk |
| Surface potential ($\Omega_1=\Omega_2$) | 3.334242 ± 0.01740 | 3.32426 ± 0.00638 | 3.330 ± 0.009 | 2.8524 ± 0.0051 |
| Overcontact = fill out | 2.04 % | 9.5 % | 9.3 % | 9.1 % |
| Bolometric albedo ($A_1$)<br>Bolometric albedo ($A_2$) | 0.500 (fixed)<br>0.500 (fixed) | 0.500 (fixed)<br>1.306 ± 0.100 | 0.500 (fixed)<br>0.500 (fixed) | 0.500 (fixed)<br>0.500 (fixed) |
| Gravity exponents ($\alpha_1=\alpha_2$) | 0.320 (fixed) | 0.320 (fixed) | 0.320 (fixed) | 0.320 (fixed) |
| Angular Rotation ($F_1=F_2$) | 1.000 | 1.000 | 1.000 (fixed) | 1.000 (fixed) |
| Volume 1<br>Volume 2 | 0.294384<br>0.220758 | 0.285940<br>0.199719 | 0.284945<br>0.200048 | 0.371055<br>0.144515 |
| Mass Ratio (q) | 0.8057 ± 0.0134 | 0.770 (fixed) | 0.773 ± 0.004 | 0.5020 ± 0.0016 |
| Limb Darkening $x_1$<br>Limb Darkening $x_2$ | 0.787 ± 0.066<br>0.746 ± 0.057 | 0.587 (fixed)<br>0.615 ± 0.016 | 0.600 ± 0.07<br>0.600 ± 0.07 | 0.640<br>0.640 |
| $L_1 / (L_1+ L_2)$<br>$L_2 / (L_1+ L_2) = 1- L_1$ | 0.551 ± 0.007<br>0.449 (calculated) | 0.563 ± 0.002<br>0.437(calculated) | 0.574 ± 0.001<br>0.426 (calculated) | 0.657 ± 0.001<br>0.343 (calculated) |
| $r_1$(back)<br>$r_2$(back) | 0.44458 ± 0.00625<br>0.40462 ± 0.00734 | 0.43692 ± 0.00157<br>0.39088 ± 0.00165 | 0.43783<br>0.39253 | 0.4751 ± 0.0015<br>0.3541 ± 0.0017 |
| $r_1$(side)<br>$r_2$(side) | 0.40910 ± 0.00401<br>0.36654 ± 0.00437 | 0.40464 ± 0.00113<br>0.35638 ± 0.00111 | 0.40518<br>0.35761 | 0.4448 ± 0.0011<br>0.3183 ± 0.0011 |
| $r_1$(pole)<br>$r_2$(pole) | 0.38628 ± 0.00308<br>0.34807 ± 0.00344 | 0.38300 ± 0.00090<br>0.33949 ± 0.00092 | 0.3834 0<br>0.34055 | 0.4184 ± 0.0019<br>0.3046 ± 0.0019 |
| $r_1$(point)<br>$r_2$(point) | 0.52223 ± 0.00000<br>0.47778 ± 0.00000 | 0.52687 ± 0.00000<br>0.47313 ± 0.00000 | 0.52647<br>0.47353 | 0.570351<br>0.429649 |
| Mean Radius 1<br>Mean Radius 2 | 0.41486<br>0.37711 | 0.40927<br>0.36336 | 0.40880<br>0.36357 | 0.44644<br>0.326047 |
| Surface Area 1<br>Surface Area 2 | 2.17578 ± 0.0000<br>1.81057 ± 0.0000 | 2.13748<br>1.68385 | 2.13086<br>1.68415 | 2.539807<br>1.365364 |
| **Spot of star 1**<br>Co-Latitude<br>Longitude<br>Spot Radius<br>Temp Factor | No Spot | 10°<br>135.71° ± 11.78°<br>15°<br>0.700 | No Spot | No Spot |
| $M_1$ & $M_2$ (in Solar unit) | 1.104 & 0.889 | 1.126 & 0.867 | 1.124 & 0.869 | 1.07 & 0.54 |
| $R_1$ & $R_2$ (in Solar unit) | 1.060 & 0.960 | 1.040 & 0.930 | 1.040 & 0.930 | 1.05 & 0.77 |
| $M_1$ Bol. & $M_2$ Bol. | 4.570 & 4.800 | 4.610 & 4.890 | 4.600 & 4.920 | |
| $\Sigma\ \omega\ (O-C)^2$ | 0.007425 | 0.021715 | | |

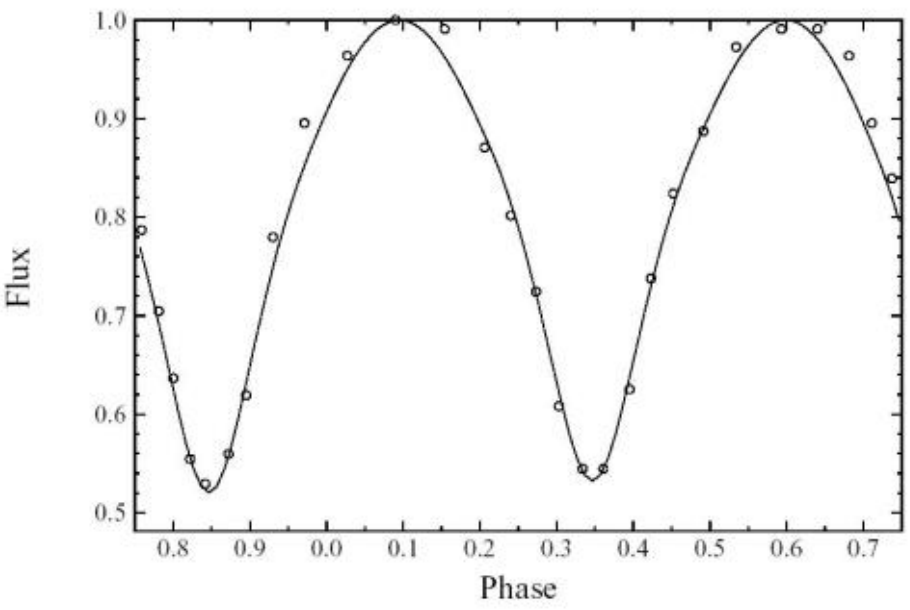

**Fig. 3: AH Tau fitted curve, 1950**

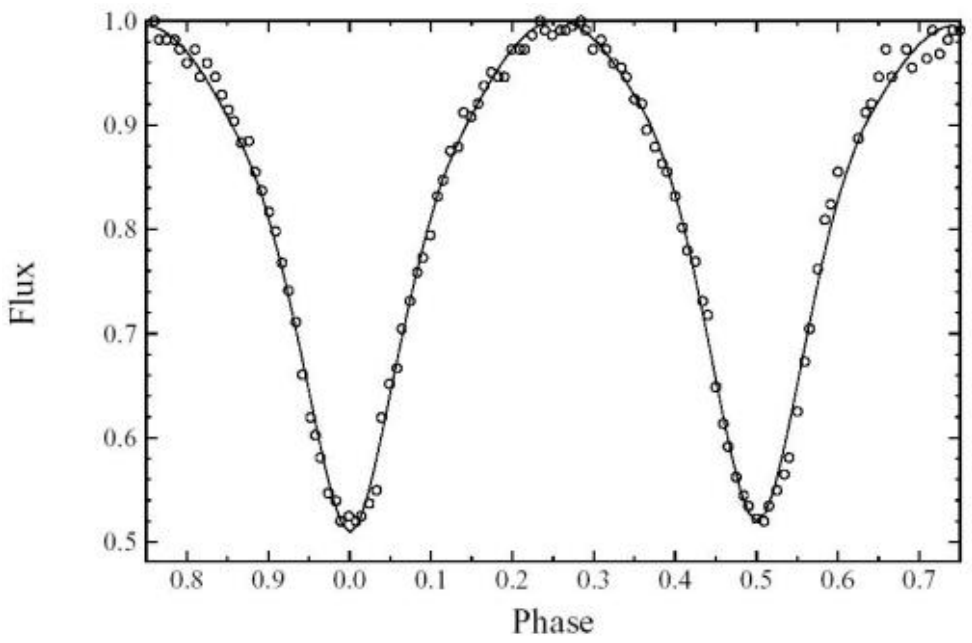

**Fig. 4: AH Tau fitted curve, 2002**

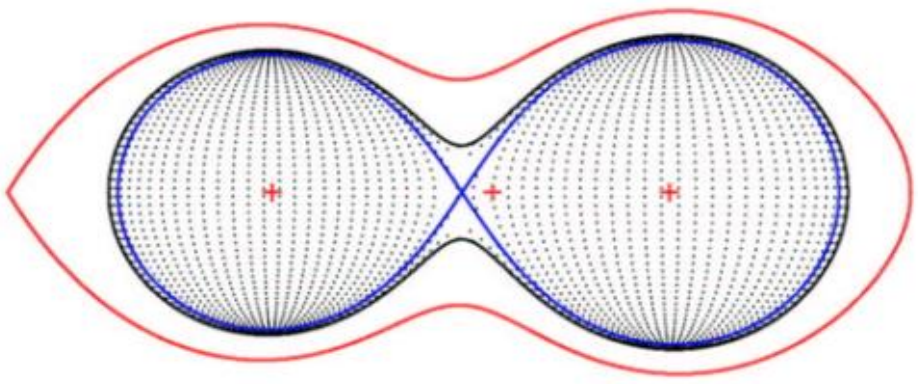

**Fig. 5: Roche Geometry of AH Tau Omega 1 = Omega = 3.334242 Overcontact = 0.203972 % B_band of 1950 Observation**

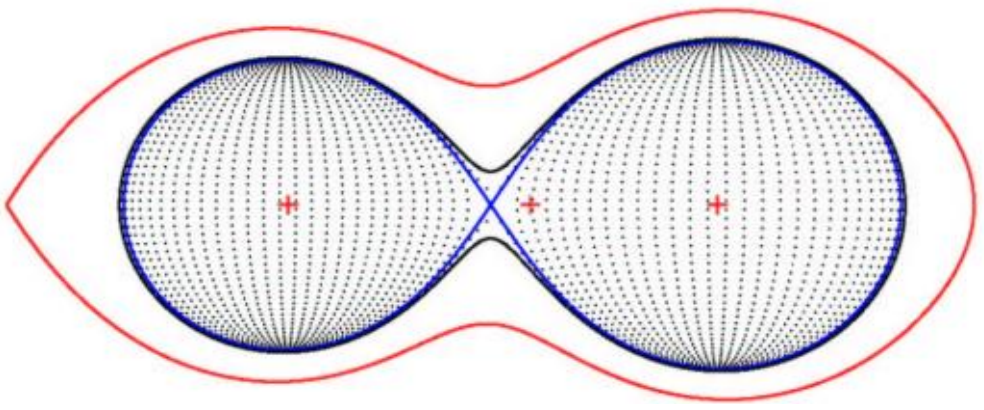

**Fig. 6: Roche Geometry of AH Tau Omega 1 = Omega 2 = 3.324262 Overcontact = 0.094456% V_band of 2002 Observation**

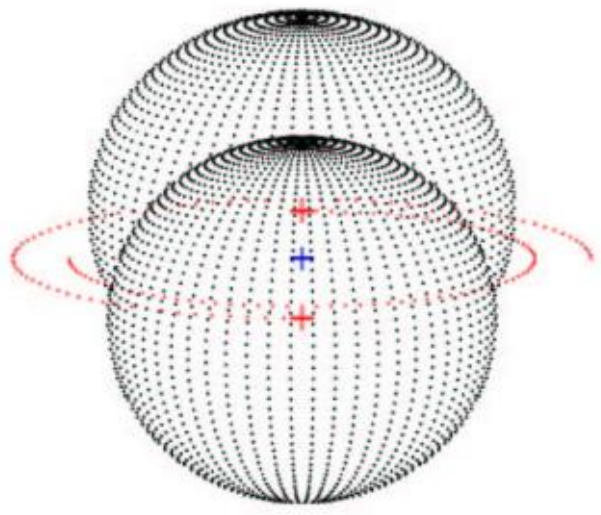

**Phase= 0.5**

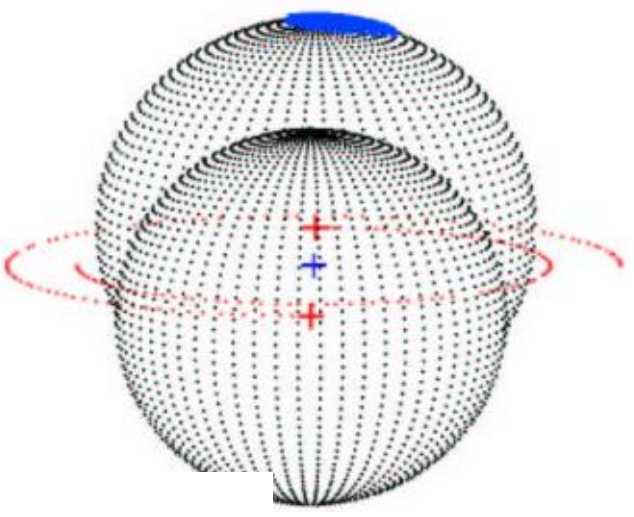

**Phase= 0.5**

**Fig. 7: : 3D modes of AH Tau for Bband of 1950 observations.**

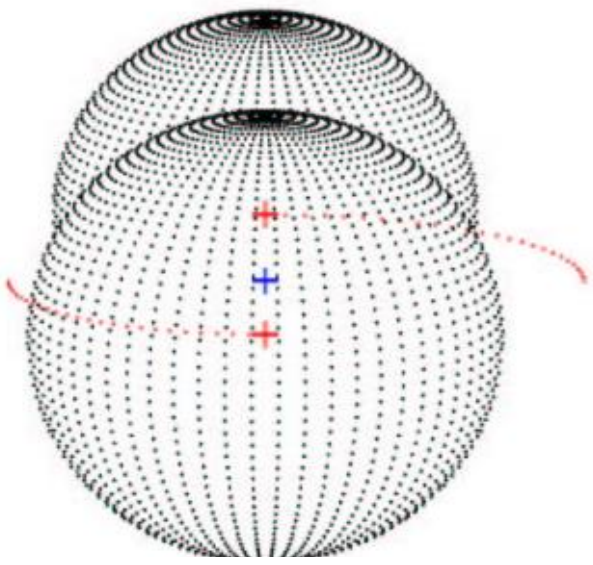

**Phase= 0.0**

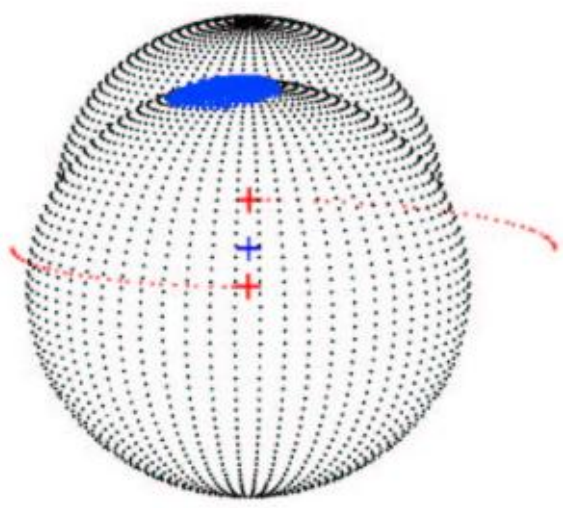

**Phase= 0.5**

**Fig. 8 : 3D modes of AH Tau for Vband 2002 observations**

**Table 2: Light Curves change of AH Tau.**

| Year | Filter | Amplitude | | Depth Difference (Pri. – Sec.) | Reference |
|---|---|---|---|---|---|
| | | Pri. Min. | Sec. Min | | |
| 1943 | Photography | 0.68 | 0.65 | 0.03 | Binnendijk, 1950 |
| 1973 | B | 0.80 | 0.73 | 0.07 | Magoloshvili, et al. 1980 |
| | V | 0.69 | 0.62 | 0.07 | |
| 1986 | B | 0.75 | 0.66 | 0.09 | Liu, et al. 1991 |
| | V | 0.71 | 0.64 | 0.07 | |
| 2002 | V | 0.71 | 0.71 | 0.00 | Nakajima, 2002 |
| 2003 | I | 0.47 | 0.43 | 0.04 | Bybath, et al. 2004 |
| | R | 0.47 | 0.43 | 0.04 | |
| | V | 0.49 | 0.45 | 0.04 | |

**Table 3: Comparison Between of the Subtype A and W of the Contact Binary System AH TAURI**

| | Property | A | | W | | Remarks |
|---|---|---|---|---|---|---|
| **1** | Spectral Type | Earlier (A-F) | | Later (G-K) | √ | Differences Slightly Marked. |
| **2** | Luminosity ( L1) | Higher | √ | Lower | | Differences Slightly Marked. |
| **3** | Activity (Changes of Light Curve, Asymmetries of Maxima) | Moderate or Absent (Symetric) | √ | Strong or Very Strong (almost Very System) | | |
| **4** | Mass- Ratio | Small 0.08-0.54 | | Larger 0.33-0.88 | √ | Upper Limit More Certain. |
| **5** | Degree of Contact (Fill Out Factor) | Envelope Slightly Thicker Than in W-Subtype | √ | Shallow Envelopes | | |
| **6** | Temperature Compare | T1> T2 | √ | T1< T2 | | |
| **7** | Mass Compare | m1> m2 | √ | m1< m2 | | |
| **8** | Light Curve Shape | • Transit at Primary Minima.<br>• Deeper Contact.<br>• Langer Period. | √ | • Occultation at Primary Minima.<br>• Shallow Contact.<br>• Short Period. | | From the most published light curves |
| | | • Evolved | | • Less Certain Unstable. | | |

## DISCUSSIONS AND CONCLUSION

The system AH Tau. has been analyzed previously by WD code by Liu et al (1991) and by Byboth et al (2004). In the present work the system has been analyzed too by WD code for the Binnendijk's observation (1950) and the Nakajima's observation (2002), the parameters of the system for all pervious analysis have been represented in Table 1.
The best fit for the two observational sets in B bands for Binnendijk (1950) and V band for Nakajima (2002) are shown in figures 3 and 4 respectively.

From the morphology examination of the light curves for the system, We can find no difference in the maxima for all light curves which have been observed, while the eclipse depths are changeable. So, the asymmetry of the system only shown in the eclipse depths. Table 2 represents the year of observations and the corresponding amplitudes for the primary and secondary eclipse for most light curves observed completely. While table 3 represents the schemes for the classification of the subtype A- and W- of the system according to Rucinski (1974) and Binnendijk (1970).
The system configuration at different phases (0.0, 0.5) is shown in figure 7 for B filter of 1950, and is shown with spot position in figure 8 for V filter of 2002 observations.

From the light curves shape, the primary minimal (minimum) may almost in transit, while the two maxima um have the same level. So, the categorization of the eclipsing binary AH Tau. may like to be an A-subtype of W- UMa system, according to the scheme of table 3, and the schematic picture of its internal structure in figures 5 and 6 (Rucinski, 1974). Binnendijk (1950) classified the system as W-UMa, while Romano (1962) classified it to be of β Lyr type.

The asymmetry has been found in the minima for all light curves observed, as shown in table 2, except the light curve observed by Nakajima (2002). The best fit of this light curve has been found by a dark spot model on the massive component. From the analysis of the light curves, Binnendijk (1950) and the Nakajima (2002), as well as the analysis of Byboth et al (2004) and Liu, et al. (1991), it is shown that the rimary component is slightly massive than the secondary one. The surface area of the primary is larger than the secondary by a factor of about 1.2, as shown from Table 1. The luminosities of the two components are slightly different too, while the temperatures $T_1$ & $T_2$ for the two components are comparable. So the dark spot may influence the asymmetry mode.

Bookmayer (1971) indicated that the spectral type of around G5 while Binnendijk (1950), according to his photographic observation and the colour indices of the system, referred the spectral type of the system to be a G1p.

So, we can conclude that the Solar type star AH Tau. may be considered as an A- subtype W-UMa contact binary with a fill out ratio 9.5 %. The dark spot on the massive component may affect the asymmetry of the eclipse depth of the light curves with the temperature spot factor of 0.7. The sudden symmetry in the eclipse depth of the light curve for the contact binary AH Tau. system found from the Nakajima's (2002) observations, needs further studies according to the recent stellar structure model.

# REFERANCE